\documentclass[fleqn,twoside]{article}
\usepackage{espcrc2,psfig}

\newlength{\figwidth}
\newcommand{\Hm}{\hspace{-0.24cm}}
\renewcommand{\vec}[1]{\mbox{\boldmath $#1$}}

\title{$O(a)$ improved Wilson quark action on anisotropic lattice%
       \thanks{Talks presented by T. Umeda and H. Matsufuru.}}

\author{
T. Umeda\address{%
    Center for Computational Physics, University of Tsukuba,
    Tsukuba 305-8577, Japan \vspace{-0.2cm}},
H. Matsufuru\address{%
    Yukawa Institute for Theoretical Physics, Kyoto University,
    Kyoto 606-8502, Japan}
and
T.~Onogi$^{\rm b}$}

\begin{document}

\begin{abstract}
The $O(a)$ improved Wilson quark action on the anisotropic lattice
is investigated.
We carry out numerical simulations in the quenched approximation
at three values of lattice spacing ($a_{\sigma}^{-1}=1$--$2$ GeV)
with the anisotropy $\xi=a_{\sigma}/a_{\tau}=4$,
where $a_{\sigma}$ and $a_{\tau}$ are the spatial and the temporal
lattice spacings, respectively.
Using the dispersion relation of mesons,
the bare anisotropy $\gamma_F$ in the quark action is
numerically tuned below the charm quark mass region
with the statistical accuracy of 1 \% level.
The systematic uncertainties in the calibration are examined and
found to be under control in the continuum limit.
Then we compute the light hadron masses and
find that they are consistent with the result of the UKQCD
Collaboration on the isotropic lattice.
The effect of the uncertainty in the calibration on the
hadron spectrum for physical quark masses is also found to be
under control.
\end{abstract}

\maketitle

\section{Introduction}
  \label{sec:introduction}

The anisotropic lattice, which allows a finer lattice spacing in the 
temporal direction, is expected to be useful for 
physics such as the spectroscopy of exotic states,  
finite temperature QCD and the heavy quark physics.
Since the manifest temporal-spatial axis interchange symmetry is
absent, the anisotropy parameters of the action should be tuned
by imposing the conditions with which the Lorentz invariance is
satisfied for some physical observables. A lot of effort on 
anisotropic lattices has been devoted to charmonium systems
\cite{Ume01,Kla99,Chen00,CPPACS01}, but because of the 
computational cost for calibration, the light 
quark on the anisotropic has not been explored so far.

In this report, we present our study on the $O(a)$ improved Wilson
quark action in the light quark region \cite{Aniso01b}
for the range of quark mass from the massless limit up to
around the charm quark mass on the quenched anisotropic lattices
with three lattice spacings,
$a_{\sigma}^{-1}=$1--2 GeV, at fixed renormalized anisotropy,
$\xi=a_{\sigma}/a_{\tau}=4$, where $a_{\sigma}$ and $a_{\tau}$ are
the spatial and temporal lattice spacings.
First the bare anisotropy in the quark action is numerically tuned
so that the renormalized fermionic anisotropy is equal to that
of the gauge field by imposing the relativistic dispersion relation
of mesons.
The systematic uncertainties due to the anisotropy are examined.
Then we compute the light hadron spectrum and examine how
the uncertainty in the calibration affect the spectrum at the
parameters of physical interest.
More detailed discussions were presented in \cite{Aniso01b}.

\section{Quark action and dispersion relation of free quark}
\label{sec:formulation}

The quark action is the same as the Fermilab action \cite{EKM97}
but defined on an anisotropic lattice as has been discussed 
in Ref.~\cite{Aniso01a}. In the hopping parameter form, 
\begin{eqnarray}
 S_F &=& \sum_{x,y} \bar{\psi}(x) K(x,y) \psi(y), \\
 K(x,y) \!\!\!&=&\!\!\!
 \delta_{x,y}
   - \kappa_{\tau} \left[ \ \ (1-\gamma_4)U_4(x)\delta_{x+\hat{4},y} \right.
 \nonumber \\
 & &  \hspace{1cm}
      + \left. (1+\gamma_4)U_4^{\dag}(x-\hat{4})\delta_{x-\hat{4},y} \right]
 \nonumber \\
 & & \hspace{-0.2cm}
    -  \kappa_{\sigma} {\textstyle \sum_{i}}
         \left[ \ \ (r-\gamma_i) U_i(x) \delta_{x+\hat{i},y} \right.
 \nonumber \\
 & & \hspace{1cm}
     + \left. (r+\gamma_i)U_i^{\dag}(x-\hat{i})\delta_{x-\hat{i},y} \right]
 \nonumber \\
 & & \hspace{-0.2cm}
    -  \kappa_{\sigma} c_E
             {\textstyle \sum_{i}} \sigma_{4i}F_{4i}(x)\delta_{x,y}
 \nonumber \\
 & & \hspace{-0.2cm}
    - r \kappa_{\sigma} c_B
             {\textstyle \sum_{i>j}} \sigma_{ij}F_{ij}(x)\delta_{x,y},
 \label{eq:action}
\end{eqnarray}
where $\kappa_{\sigma}$ and  $\kappa_{\tau}$ 
are the spatial and temporal hopping parameters, $r$ is the Wilson
parameter and  $c_E$ and $c_B$ are  the clover coefficients.
In principle for a given $\kappa_{\sigma}$, the 
four parameters $\kappa_{\sigma}/\kappa_{\tau}$, $r$, $c_E$ and $c_B$
should be tuned so that Lorentz symmetry holds up to 
discretization errors of $O(a^2)$.

In this work, we set the spatial Wilson parameter as $r=1/\xi$ and
the clover coefficients as the tadpole-improved tree-level
values, namely,
\begin{equation}
 r = 1/\xi, \hspace{0.5cm}
 c_E= 1/u_{\sigma} u_{\tau}^2, \hspace{0.5cm}
 c_B = 1/u_{\sigma}^3
\label{eq:cecb}
\end{equation}
and perform a nonperturbative calibration only for $\gamma_F$.
The tadpole improvement \cite{LM93} is achieved
by rescaling the link variable as
$U_i(x) \rightarrow U_i(x)/u_{\sigma}$ and  $U_4(x) \rightarrow
U_4(x)/u_{\tau}$, with the mean-field values of the spatial 
and temporal link variables, $u_{\sigma}$ and $u_{\tau}$,
respectively.
This is equivalent to redefining the
hopping parameters with the tadpole-improved ones (with tilde)
through $\kappa_{\sigma} = \tilde{\kappa}_{\sigma}/u_{\sigma}$
and $\kappa_{\tau} = \tilde{\kappa}_{\tau}/u_{\tau}$.
We define the anisotropy parameter $\gamma_F$ as
$\gamma_F \equiv \tilde{\kappa}_{\tau}/\tilde{\kappa}_{\sigma}$.

For later convenience, we also introduce $\kappa$ 
\begin{eqnarray}
\frac{1}{\kappa} \equiv \frac{1}{\tilde{\kappa}_{\sigma}}
     - 2(\gamma_F+3r-4)
\hspace{0.3cm}  = 2(m_{0\sigma}+4) ,
 \label{eq:kappa}
\end{eqnarray}
where $m_{0\sigma}$ is the bare quark mass in spatial lattice units.

The free quark propagator for the action (\ref{eq:action}) satisfies
the dispersion relation
\begin{equation}
\cosh E(\vec{p}) = 1 + \frac{\vec{\bar{p}}^2
            + (m_0 + \frac{1}{2} \frac{r}{\gamma_F} \vec{\hat{p}}^2 )^2}
            { 2 (1+m_0 + \frac{1}{2} \frac{r}{\gamma_F} \vec{\hat{p}}^2 ) },
\label{eq:dispersion}
\end{equation}
where $\bar{p}_i = \frac{1}{\gamma_F}\sin p_i$,
$\hat{p}_i=2\sin(p_i/2)$, and $m_0=m_{0\sigma}/\gamma_F$ is the
bare quark mass in temporal lattice units.
The rest mass $M_1$ and the kinetic mass $M_2$ are defined as
  $M_1 \equiv E(\vec{0})$ and
  $1/M_2 \equiv \xi^2 d^2 E / dp_i^{\,2} |_{\vec{p}=0}$.
One can tune the bare anisotropy parameter $\gamma_F$
so that the rest and kinetic masses give the same values \cite{EKM97}.
For small $m_0$ and with $r=1/\xi$,
this leads the expansion of $\gamma_F$ in $m_0$ as
\begin{equation}
\gamma_F^{-1}\
 = \xi^{-1} \left[ 1 + \mbox{$\frac{1}{3}$} m_0^2 \right] .
\label{eq:gamma_F_in_m_0}
\end{equation}
The $m_0$ dependence starts with the quadratic term for $r=1/\xi$;
therefore the dependence on the quark mass is small for sufficiently
small $m_0$.
For example, let us consider the case of $a_{\tau}^{-1}=4$ GeV,
which corresponds to our coarsest lattice in the simulation.
The charm quark mass corresponds to $m_0\simeq 0.3$ and at this
value $\gamma_F$ is different from $\xi$ by only 3\%.
Up to this quark mass region, one can expect that the difference of
$\gamma_F$ from $\xi$ will also be small in the numerical simulation.

\begin{figure}[tb]
\vspace*{-0.2cm}
\psfig{file=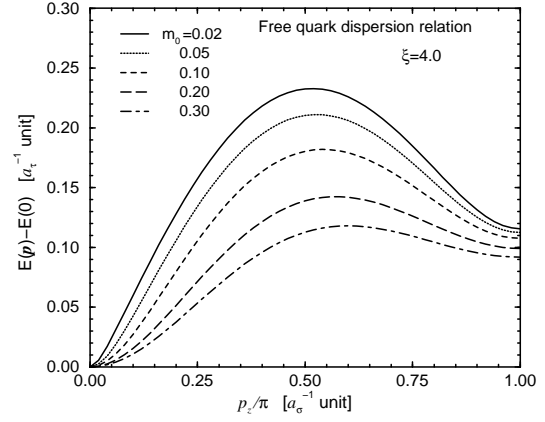,width=\figwidth}
\vspace{-1.2cm}
\caption{Dispersion relation of the free quark for $\xi=4$.}
\label{fig:dispersion}
\vspace{-0.4cm}
\end{figure}

With our choice, $r=1/\xi$, the action (\ref{eq:action}) leads to a
smaller spatial Wilson term for a larger anisotropy $\xi$.
The question is whether the contribution of the doubler eliminated
by the Wilson term becomes significant for practical value of $\xi$.
Figure~\ref{fig:dispersion} shows the dispersion relation
(\ref{eq:dispersion}) for several values of $m_0$
in the case of $\xi=4$.
Towards the edge of Brillouin zone, the relative energy
$E(\vec{p})-E(0)$ rapidly decreases to $\simeq 0.1$
in temporal lattice units.
On the other hand, typical energy scale of quarks inside hadrons
are about $\Lambda_{QCD}\simeq 200$ MeV for light and heavy-light
hadrons.
For our coarsest lattice, $a_{\tau}^{-1}\simeq 4.0$ GeV leads
to the doubler's relative energy of about 400 MeV.
This seems large enough for eliminating the naive excited state
contamination due to the doublers from the ground state signals,
although more nontrivial doubler effect such as the shift of energy 
through the mixing of the ground states and the doubler states is not excluded.
Our finest lattice has $a_{\tau}^{-1}\simeq 8$ GeV
which would be sufficiently large to avoid the doubler effect.
In the case of heavy quarkonium, typical kinetic energy scale is
$mv^2$, where $v$ is the quark velocity.
For charmonium, $v^2 \sim 0.3$ and typical energy scale is about
$300$ MeV.
Since this is not sufficiently less than the doubler's relative
energy on the lattice of $a_{\tau}^{-1}\simeq 4.0$ GeV,
finer lattices should be used for the charmonium system.

\section{Calibration procedures}
\label{sec:calibration}

On an anisotropic lattice, one must tune the parameters so that the
anisotropy of quark field, $\xi_F$, equals that of the gauge
field, $\xi_G$:
\begin{equation}
 \xi_F(\beta,\gamma_G;\kappa,\gamma_F)
 = \xi_G(\beta,\gamma_G;\kappa,\gamma_F)
= \xi .
\label{eq:xi}
\end{equation} 
Although $\xi_G$ and $\xi_F$ are in general functions of both of
gauge parameters ($\beta$, $\gamma_G$) and quark parameters
($\kappa$, $\gamma_F$),
on a quenched lattice one can determine $\xi_G=\xi$ independently of
$\kappa$ and $\gamma_F$, and then tune $\gamma_F$ so that a certain
observable satisfies the condition (\ref{eq:xi}).
In this work, we define $\xi_F$ through the relativistic dispersion
relation of meson,
\begin{equation}
E^2 (\vec{p}) = m^2 + \vec{p}^2 / \xi_F^2 + O[(\vec{p}^2)^2],
\label{eq:DR1}
\end{equation}
for calibration.
In the above expression, the energy and mass $E$ and $m$ 
are in temporal lattice units while the momentum $\vec{p}$ is
in spatial lattice units.
$\xi_F$ converts the momentum into that in temporal lattice units.
For finite lattice spacings, the above dispersion relation only holds
up to the $O[(\vec{p}^2)^2]$ correction term.
In the continuum limit, this higher order term in $a$ would
vanish and the relativistic dispersion relation would be restored.

We fit our numerical data for $E^2$ to the form Eq.~(\ref{eq:DR1})
and obtain the value of $\xi_F$ for each input value of bare anisotropy
$\gamma_F$.
Then we linearly interpolate $\xi_F$ in terms of $\gamma_F$
and find $\gamma_F^*$, the value of $\gamma_F$ for which 
$\xi_F=\xi$ holds.

In order to estimate the systematic errors we also 
use the dispersion relation that corresponds to 
the lattice Klein-Gordon action \cite{Ume01},
\begin{equation}
\cosh E(\vec{p}) - \cosh E(\vec{p}=0)
   =  \vec{\hat{p}}^2 / 2\xi_{KG}^2 .
\label{eq:DR2}
\end{equation}
The difference between these two calibration conditions 
shows the typical size of the lattice discretization errors.
Expanding this expression in $a$, 
$\xi_{KG}$ is related to $\xi_F$ as
\begin{equation}
  \xi_{KG} = \xi_F [ 1 - m^2/12 +O(a^4) ].
\label{eq:DR3}
\end{equation}
The same input $\gamma_F$ gives a smaller value for
$\xi_{KG}$ than $\xi_F$, and therefore the tuned bare anisotropy
$\gamma_F^{*}$ results in a larger value in the former case.

\section{Numerical Results of Calibration}
\label{sec:numerical}

\subsection{Simulation Parameters}

Numerical simulations are performed on three quenched lattices
with the Wilson plaquette action at $\beta=5.75$, $5.95$, and $6.10$
and with the renormalized anisotropy $\xi=4$.
For the values of bare anisotropy of gauge field $\gamma_G$,
we adopt the relation of $\gamma_G$ to $\xi$ numerically
determined by Klassen in one percent accuracy \cite{Kla98}.
Table~\ref{tab:parameters} summarizes the simulation parameters.
The configurations are fixed to the Coulomb gauge,
which is particularly useful for the smearing of hadron operators.

The lattice cutoffs and the mean-field values of link variables
in Table~\ref{tab:parameters}
are determined on the smaller lattices with half size in temporal
extent for $\beta=5.75$, $5.95$, 
and otherwise with the same parameters,
while at $\beta=6.10$ the lattice size is $16^3\times 64$.
The lattice cutoffs are determined from the hadronic radius $r_0$
proposed in \cite{Som94}, by setting the physical value of
$r_0$ as $r_0^{-1}=395$ MeV.
The mean-field values, $u_{\sigma}$ and $u_{\tau}$, are obtained
as the average of the link variables in the Landau gauge,
where the mean-field values are used
self-consistently in the fixing condition \cite{Ume01}.

\begin{table}[tb]
\caption{
Lattice parameters.
The statistical uncertainty of  $u_{\tau}$ is less than the last
digit.}
\begin{center}
\small
\begin{tabular}{cccccc}
\hline\hline
$\beta$ & $\gamma_G$ & size &
\hspace{-0.3cm} $a_{\sigma}^{-1}$ [GeV] \hspace{-0.3cm} &
\Hm  $u_{\sigma}$ & \Hm $u_{\tau}$  \\
\hline
\hspace*{-0.2cm} 5.75 & \Hm 3.072~ & \Hm $12^3\!\times\! ~96$ &
 \Hm 1.100( 6) & \Hm 0.7620(2) & \Hm 0.9871 \hspace{-0.3cm} \\
\hspace*{-0.2cm} 5.95 & \Hm 3.1586 & \Hm $16^3\!\times\! 128$ &
 \Hm 1.623( 9) & \Hm 0.7917(1) & \Hm 0.9891 \hspace{-0.3cm} \\
\hspace*{-0.2cm} 6.10 & \Hm 3.2108 & \Hm $20^3\!\times\! 160$ &
 \Hm 2.030(13) & \Hm 0.8059(1) & \Hm 0.9901 \hspace{-0.3cm} \\
\hline\hline
\end{tabular}
\end{center}
\vspace{-0.6cm}
\label{tab:parameters}
\end{table}

\subsection{Quark field calibration}
\label{subsec:calbration_summary}

The calibration of the bare anisotropy $\gamma_F$ in the 
quark action is performed as described in the last section.
The tuned bare anisotropy parameter $\gamma_F^*$, at which $\xi_F=\xi$
holds, is determined in the region from strange to charm quark masses
using the dispersion relation of the pseudoscalar and vector mesons.
The meson energies are extracted from the meson correlators
with momenta $\vec{p}=\vec{n}(2\pi/L)$, where $L$ is the spatial
lattice extent and
$\vec{n}=(0,0,0)$, $(1,0,0)$, $(1,1,0)$, $(1,1,1)$, and $(2,0,0)$,
where we take averages over rotationally equivalent momenta. 
The energies are fitted to both linear and quadratic forms in $\vec{p}^2$
in order to obtain $\xi_F$ in each channel, where
the linear fit always uses three lowest momentum states.
We adopt the result of quadratic fit at $\beta=6.10$, except
for a few lightest quark cases in which the statistical fluctuation
in large momentum states is severely large.
At $\beta=5.75$ and $5.95$, linear fit is used in the whole quark mass
region, since quadratic fits which include higher momentum states 
suffer from larger discretization errors.

\begin{figure}[tb]
\vspace*{-0.2cm}
\psfig{file=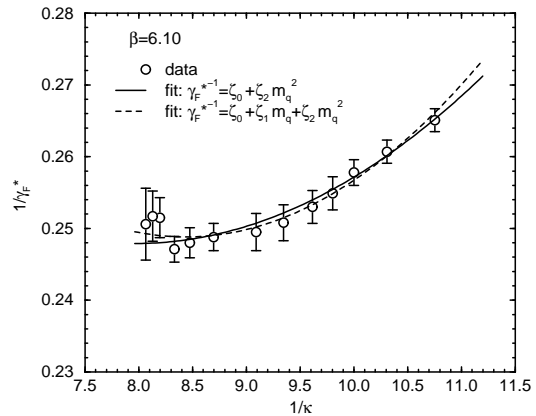,width=\figwidth}
\vspace{-1.2cm}
\caption{$1/\gamma_F^*$ vs $1/\kappa$ at $\beta=6.10$.
The solid line shows the fit linear in $m_q^2$
while the dashed line represents the fit quadratic in $m_q$.}
\label{fig:zeta}
\vspace{-0.4cm}
\end{figure}

We measure $\xi_F$  for two to four different values of $\gamma_F$.
By linear interpolation we find the value of $\gamma_F^*$ 
in each channel of PS and V mesons.
We take the average of 
$\gamma_F^*{(PS)}$ and $\gamma_F^*{(V)}$ as the central value
and use the difference to estimate one of the systematic errors.
At each $\kappa$, $\gamma_F^*$ is obtained within 1\% statistical
error, except for the lightest quark mass region.
Figure~\ref{fig:zeta} shows the $\kappa$ dependence of $\gamma_F^*$
at $\beta=6.10$.
Similar behaviors are observed on other two lattices.
To determine $\gamma_F^*$ precisely in the light quark mass region
and to extrapolate to the chiral limit, we fit $\gamma_F^*$ to
the form
\begin{equation}
 \frac{1}{\gamma_F^*}(m_q) = \zeta_0 + \zeta_2 m_q^2,
\end{equation}
where the quark mass in temporal units $m_q$ is defined  as 
$m_q = \frac{1}{2 \xi} ( \frac{1}{\kappa} 
    - \frac{1}{\kappa_c} )$.
The critical hopping parameter $\kappa_c$ are determined by a linear
extrapolation of PS meson mass squared in terms of $1/\kappa$
with two largest $\kappa$'s.
The result of the fits is listed in Table~\ref{tab:fit_gamma_F},
and represented by the solid line in Figure~\ref{fig:zeta}.
The linear fit in $m_q^2$ seems quite successful, and $\gamma_F^*$ at
the chiral limit is close to the tree-level value, $\xi$.
The statistical uncertainty in $\gamma_F^*$ is estimated to be
of the order of 1\% for the whole quark mass region.
We also carry out a quadratic fit in $m_q$, as shown in
Fig.~\ref{fig:zeta} by the dashed line, and find that the difference
in $\gamma_F^*(m_q=0)$ from the linear fit in $m_q^2$ is at most 1\%.
This difference should be considered as the systematic uncertainty
in the chiral extrapolation.

\begin{table}[tb]
\caption{Fit results for $\gamma_F^*$.}
\begin{center}
\begin{tabular}{cccc}
\hline\hline
$\beta$ & 
$\zeta_0$ & $\zeta_2$ & $\kappa_c$  \\
\hline
5.75& 0.2558( 9)& 0.230(12) & 0.12640(5)\\
5.95& 0.2490( 8)& 0.189(15) & 0.12592(8)\\
6.10& 0.2479( 9)& 0.143(14) & 0.12558(9)\\
\hline\hline
\end{tabular}
\end{center}
\label{tab:fit_gamma_F}
\vspace{-0.6cm}
\end{table}

\subsection{Uncertainties in calibration} 

To examine the uncertainty in the calibration, 
the following analyses are also carried out.

(i). We measure the difference between the $\gamma_F^*$'s for
the pseudoscalar and vector mesons, which signals the $O(\alpha a)$
systematic error.
They tend to decrease towards smaller lattice
spacing, and is already consistent with zero at $\beta=5.95$.

(ii). To estimate the size of $O(a^2)$ systematic uncertainty,
$\gamma_F^*$ is also determined using the lattice Klein-Gordon type
dispersion relation, Eq.~(\ref{eq:DR2}).
The results with two dispersion relations come closer to
each other by decreasing the lattice spacing.
The sizes of this uncertainty in the light quark mass region
are 3\%, 2\%, and 1\% for $\beta=5.75$, $5.85$, and $6.10$, respectively.
The behavior in the large quark mass region is consistent with
what is expected from Eq.~(\ref{eq:DR3}).

(iii). We obtain the response of meson masses to a change of
$\gamma_F$ at the fixed $\kappa$.
The effect of uncertainty in $\gamma_F^*$ on the meson masses is less than
1\%, if $\gamma_F^*$ is determined at this accuracy.
This result is applicable to the relatively heavier quark mass region, 
such as $m_s < m_q$, and therefore in this region, the errors in the 
calibration are under control.
The light quark mass region is examined in the next section.

\section{Light hadron spectroscopy}
\label{sec:spectroscopy}

\subsection{Calculation of hadron spectrum}

Taking the central value of $\gamma_F=\gamma_F^*$,
the light hadron masses are computed in 
the strange quark mass region $m_s < m_q < 2 m_s$
on the same lattices used in the calibration, while with 
smaller statistics.
The parameters are listed in Table~\ref{tab:parameters2}.
In this region, we regard that $m_q$ is sufficiently small
and adopt the value of $\gamma_F^*$ in the massless limit.

\begin{table}[tb]
\caption{
Quark parameters used in the hadron spectroscopy.
The numbers of configurations are 200 at $\beta=5.75$
and 100 at $\beta=5.95$ and $6.10$.}
\begin{center}
\begin{tabular}{ccc}
\hline\hline
$\beta$ & $\gamma_F$ & values of $\kappa$ \\
\hline
5.75& 3.909 & 0.1240, 0.1230, 0.1220, 0.1210 \\
5.95& 4.016 & 0.1245, 0.1240, 0.1235, 0.1230 \\
6.10& 4.034 & 0.1245, 0.1240, 0.1235, 0.1230 \\
\hline\hline
\end{tabular}
\end{center}
\label{tab:parameters2}
\vspace{-0.6cm}
\end{table}

The quark propagators are smeared at the source with Gaussian
smearing function with the deviation $\simeq 0.4$ fm,
in the Coulomb gauge.
For baryons, two of the quarks are set to have degenerate masses.

In order to avoid the ambiguities in the definition 
of the quark mass, we extrapolate the hadron masses
to the chiral limit in terms of the pseudoscalar
meson mass squared, instead of $1/\kappa$.
We assume the relation
\begin{equation}
 m_{PS}^2(m_1, m_2) = B (m_1 + m_2);
\end{equation}
so that 
$m_{PS}^2 = 2 B m_1$ holds for the degenerate quark masses, $m_1=m_2$.
Instead of $m_i$ (i=1,2), one can use $m_{PS}(m_i,m_i)^2$
as the variable in the chiral extrapolation.
For vector mesons and octet and decuplet baryons,
we also use the linear relations.
The linear fit looks successful.

\subsection{Spectrum at physical quark masses}

We compare our hadron spectrum at the physical quark masses
with the result in the continuum limit by UKQCD Collaboration
on an isotropic lattice \cite{UKQCD00}.
We do not extrapolate our data to the continuum limit 
for lack of the number of different $\beta$'s as well as the
statistical accuracy.
We adopt two different inputs to set the scale:
one is hadronic radius $r_0$ and the other is the $K^*$ meson mass, 
which were also adopted in \cite{UKQCD00}.
From each of these scales the physical $u$, $d$ and $s$
quark masses are determined respectively.
(We do not distinguish the $u$ and $d$ quark masses.)

\begin{figure}[tb]
\vspace*{-0.2cm}
\centerline{
\hspace*{-2.2cm}
\psfig{file=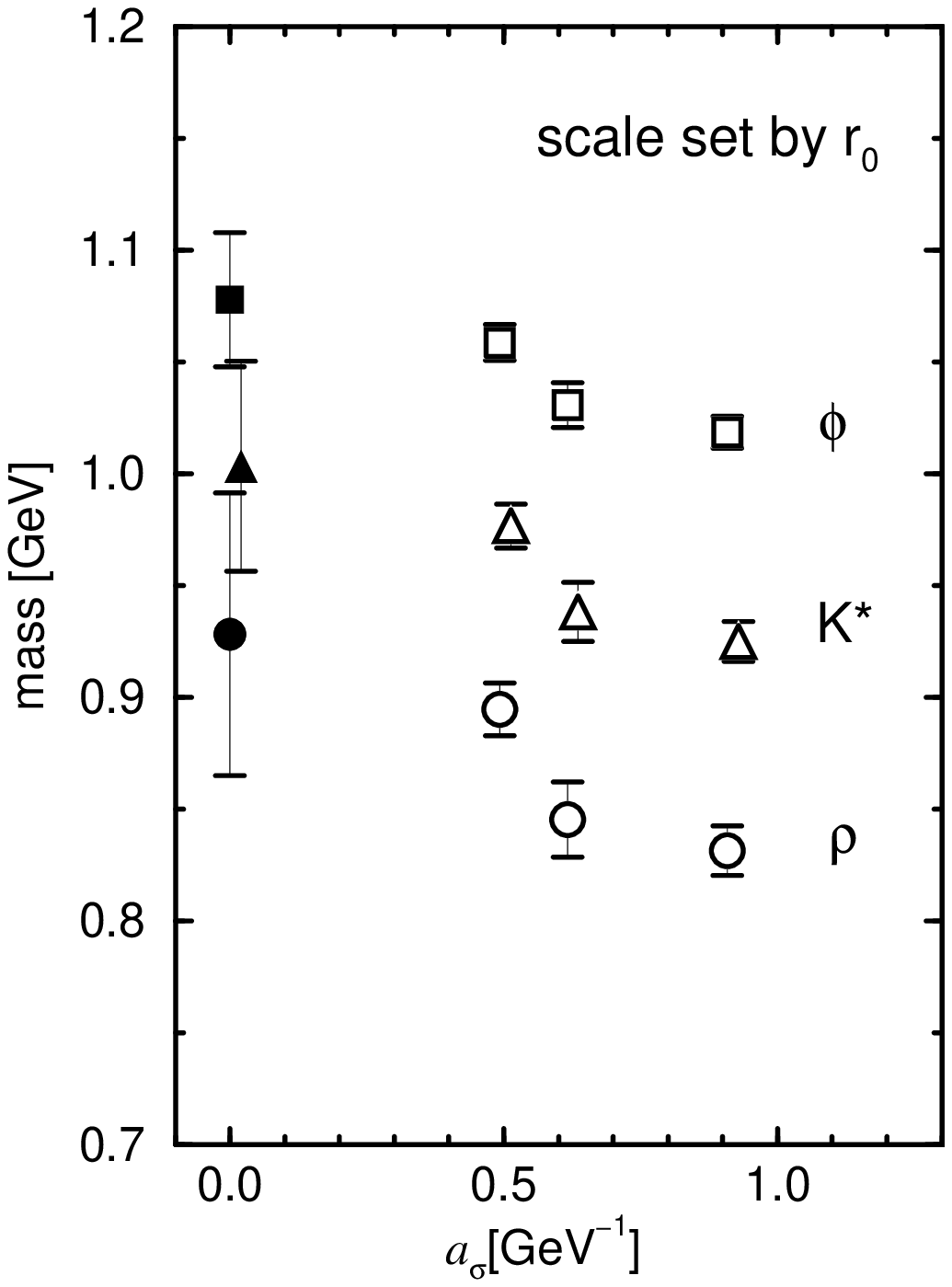,width=0.86\figwidth}
\hspace{-3.1cm}
\psfig{file=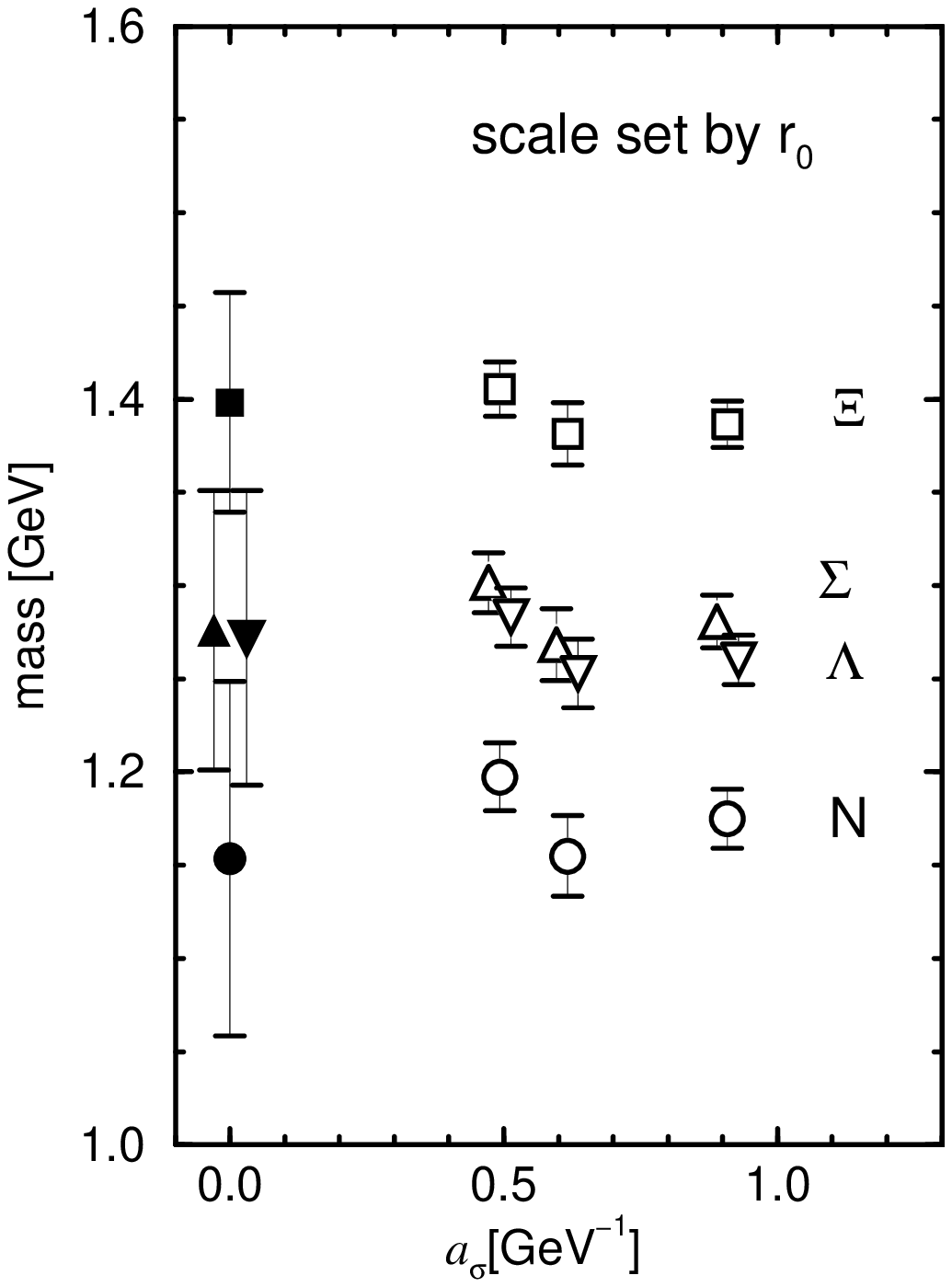,width=0.86\figwidth}}
\vspace{-0.6cm}
\centerline{
\hspace*{-2.2cm}
\psfig{file=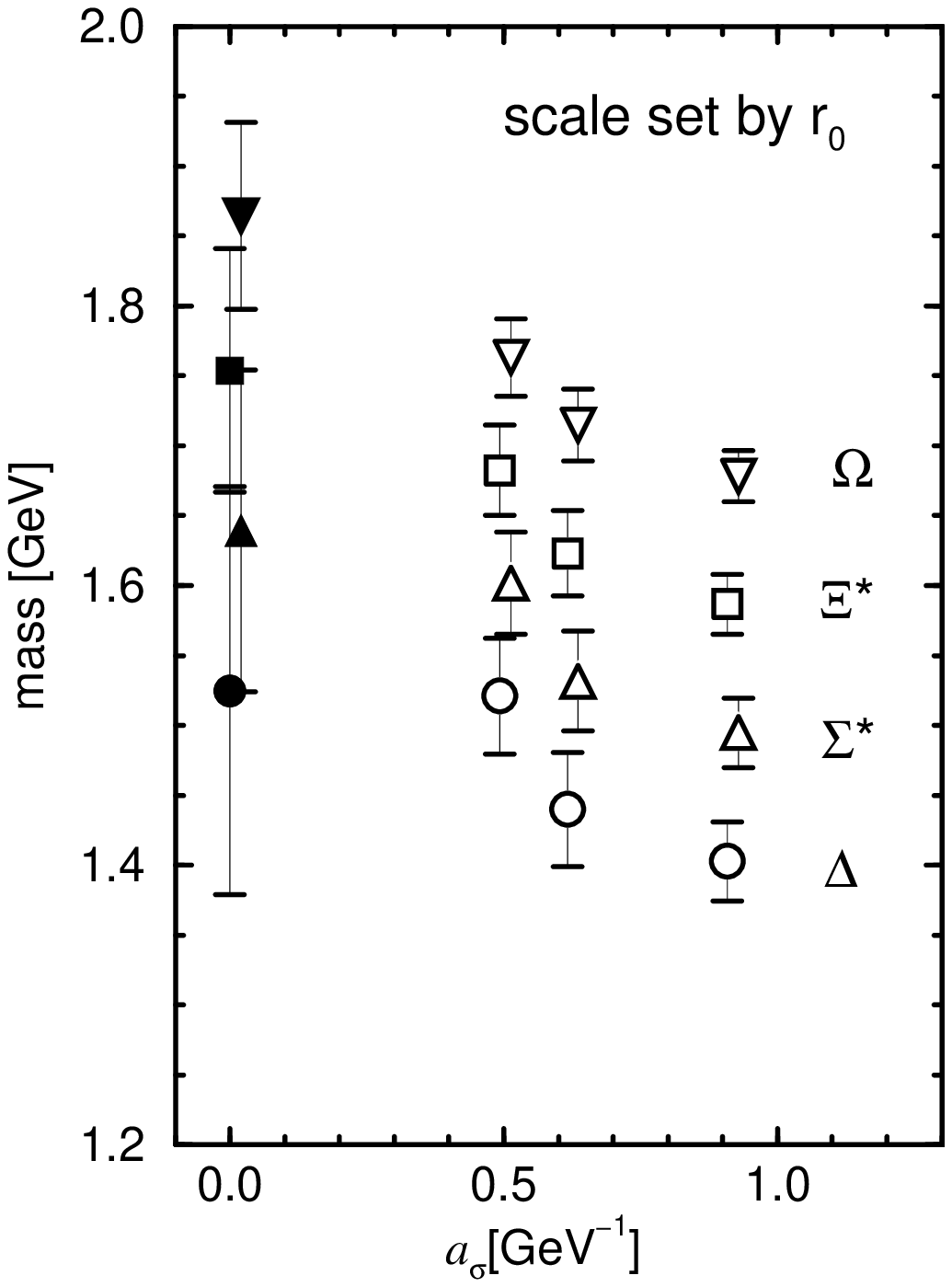,width=0.86\figwidth}
\hspace{-3.0cm}\lower.3ex
\psfig{file=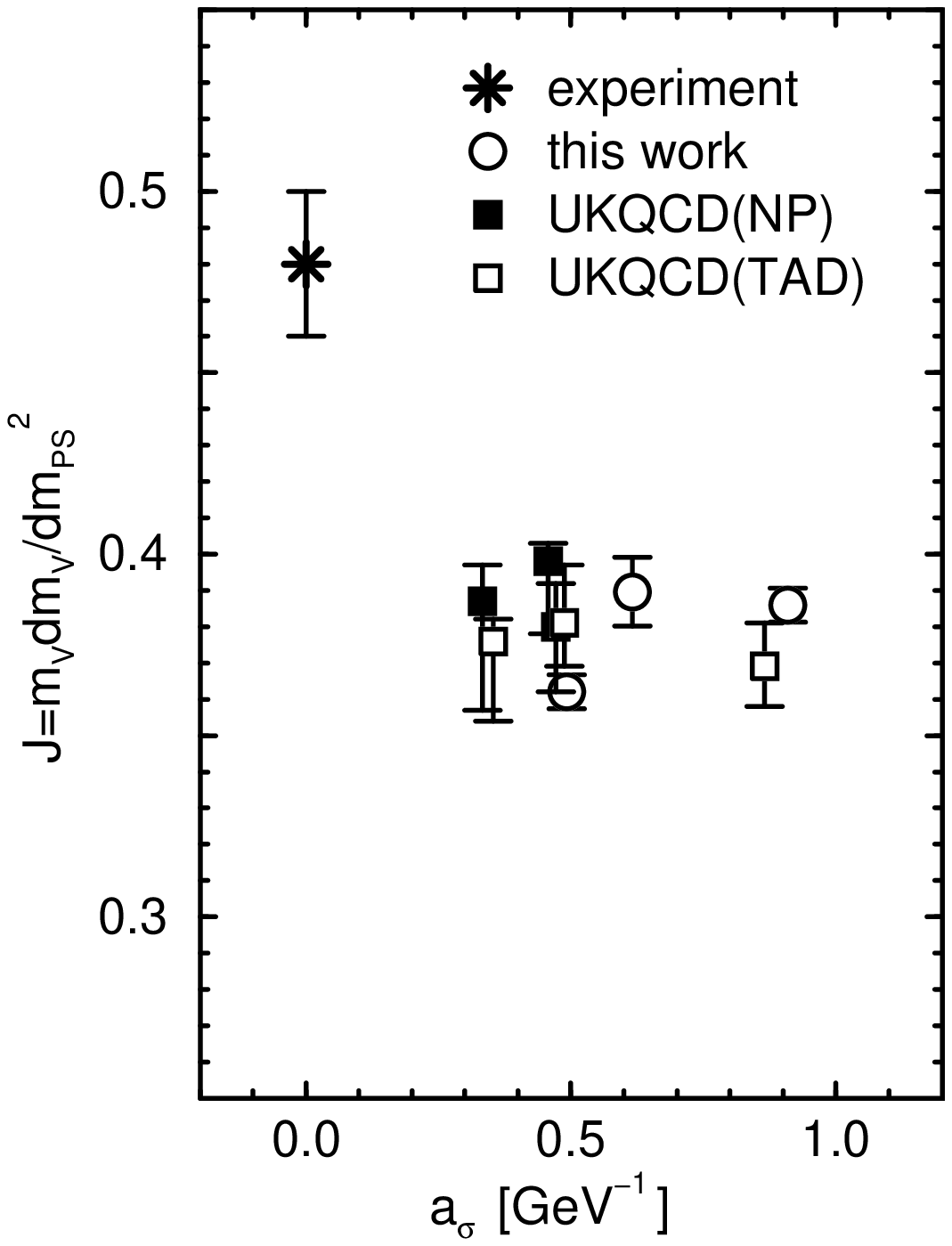,width=0.86\figwidth}}
\vspace{-1.2cm}
\caption{Hadron masses and the parameter $J$.
The $a_{\sigma}$ is set by using $r_0$.
Filled symbols are the results of UKQCD Collaboration
on the isotropic lattices \cite{UKQCD00}.}
\label{fig:masses}
\vspace{-0.4cm}
\end{figure}

While we show only the result with the scale set by $r_0$,
similar result is obtained with the scale set by $K^{*}$ meson mass.
The hadron masses extrapolated or interpolated to the physical
points are shown in Figure~\ref{fig:masses}, together with
the results of UKQCD in the continuum limit.
In our data, differences between the results at $\beta=6.10$ and
$5.95$ are rather large compared with the difference between $\beta=5.95$
and $5.75$.
This could be partially due to the different $a$ dependence of
$O(\alpha a)$ and $O(a^2)$ lattice artifacts, and also due to the
statistical fluctuation.
Our results of the hadron masses seem to approach 
the continuum results by UKQCD Collaboration on an isotropic lattice.
We also compare the parameter $J$,
\begin{equation}
 J = \left. \frac{m_V d m_V}{dm_{PS}^2}\right|_{m_V/m_{PS}=m_{K^*}/m_K},
\end{equation}
which was introduced to probe the quenching effect \cite{LM95}.
It is known that the quenched lattice simulation does not reproduce
the experimental value, $J=0.48(2)$, and gives about 20\% smaller
value.
Our results are consistent with those of UKQCD on isotropic lattices
in the quenched approximation, as shown in Fig.~\ref{fig:masses}.

\subsection{Systematic errors of the spectrum from calibration}

To estimate the effect of the uncertainty of calibration on the spectrum, 
we obtain the spectrum at the same $\kappa$'s with
slightly shifted bare anisotropy,
$\gamma_{F}' = \gamma_F^* + \delta\gamma_F$.
We set $\delta\gamma_F=0.1$, which implies about 2.5 \%
shift of bare anisotropy.
The difference between the masses with $\gamma_{F}'$
and $\gamma_F^*$ is slightly amplified toward the chiral limit.
Even for the lightest mass in each hadron species, the difference is
at most around 1\%.
This implies that the uncertainty of hadron masses at the physical
($u$,$d$) and $s$ quark masses are about half the
uncertainty in $\gamma_F$.
With the relativistic dispersion relation, $\gamma_F^*$ at $m_q=0$
has been determined at each $\beta$ within about 2\% ambiguity:
the statistical error of 1 \% and 
the systematic error of 1 \% in the form of fit.
Therefore there is 1 \% level uncertainty in the hadron spectrum
due to the uncertainty in calibration.
This feature makes the anisotropic lattice promising 
for future physical applications.

\section{Conclusion}
  \label{sec:conclusion}

We studied the $O(a)$ improved quark action
on the anisotropic lattice with anisotropy $\xi=a_{\sigma}/a_{\tau}=4$.
The bare anisotropy $\gamma_F^*$, with which $\xi_F=\xi$ holds,
is determined for the whole quark mass region below the charm quark mass,
including the chiral limit, in 1 \% statistical accuracy.
In the massless limit, there is also about 1 \% systematic uncertainty
in extrapolating $\gamma_F^*$ to $m_q=0$.
We estimate the typical sizes of $O(\alpha a)$ and $O(a^2)$ 
systematic uncertainties as to be 4\% at $\beta=5.75$ and 
smaller for larger $\beta$.
We then calculated the light hadron spectrum and found a consistent result
with previous work on an isotropic lattice by UKQCD.
The relative errors in hadron spectrum for physical quark masses
are half of those in $\gamma_F^*$.
Since finite lattice spacing errors tend to vanish as $a$ decreases,
we expect to obtain the hadron spectrum in the continuum limit
within 1\% uncertainty due to the anisotropy.
This encouraging results suggest that the anisotropic lattice 
would already be applicable to quantitative studies of
a few percent accuracy.
To achieve higher accuracy, nonperturbative tuning of 
the clover coefficients is required.

The simulation was done on a NEC SX-5 at RCNP, Osaka University
and a Hitachi SR8000 at KEK.
H.M. thank JSPS for Young Scientists for financial support.
T.O. is supported by Grants-in-Aid of the Japanese Ministry 
of Education (No. 12640279).


\begin{thebibliography}{99}

\bibitem{Aniso01b}
 H.~Matsufuru, T.~Onogi and T.~Umeda,
  hep-lat/0107001, to appear in Phys. Rev. D.

\bibitem{Ume01}
 T.~Umeda, R.~Katayama, O.~Miyamura and H.~Matsufuru,
  Int. J. Mod. Phys. A {\bf 16} (2001) 2215.

\bibitem{Kla99}
 T.~R.~Klassen,
  Nucl. Phys. B (Proc. Suppl.) {\bf 73} (1999) 918.

\bibitem{Chen00}
 P.~Chen,
  Phys. Rev. D {\bf 64} (2001) 034504.

\bibitem{CPPACS01}  
 CP-PACS Collaboration, A.~Ali Khan {\it et al.},
  Nucl. Phys. B (Proc. Suppl.)  {\bf 94} (2001) 325.

\bibitem{EKM97}
 A.~X.~El-Khadra, A.~S.~Kronfeld and P.~B.~Mackenzie,
  Phys. Rev. D {\bf 55} (1997) 3933.

\bibitem{Aniso01a}
 J.~Harada, A.~S.~Kronfeld, H.~Matsufuru, N.~Nakajima and T.~Onogi,
  Phys. Rev. D {\bf 64} (2001) 074501.

\bibitem{LM93}
 G.~P.~Lepage and P.~B.~Mackenzie,
  Phys. Rev. D {\bf 48} (1993) 2250.

\bibitem{Kla98}
 T.~R.~Klassen,
  Nucl. Phys. B {\bf 533} (1998) 557.

\bibitem{Som94}
 R.~Sommer,
  Nucl. Phys. B {\bf 411} (1994) 839.

\bibitem{UKQCD00}
 UKQCD Collaboration, K.~C.~Bowler {\it et al.},
  Phys. Rev. D {\bf 62} (2000) 054506.

\bibitem{LM95}
 UKQCD Collaboration, P.~Lacock and C.~Michael,
  Phys. Rev. D {\bf 52} (1995) 5213.

\end{thebibliography}
\end{document}